\documentclass[aps,prd,10pt,notitlepage,nofootinbib,superscriptaddress,showkeys,showpacs]{revtex4-1}
\usepackage{amsmath,amssymb,amsthm,latexsym}
\usepackage[english]{babel}
\usepackage{graphicx,color}
\usepackage{xspace}
\usepackage{graphicx}
\usepackage{pifont,dsfont}
\usepackage{marvosym}
\usepackage{slashed}
\usepackage{subfigure}

\renewcommand{\theequation}{\arabic{section}.\arabic{equation}}
\newcommand{\be}{\begin{equation}}
\newcommand{\ee}{\end{equation}}
\newcommand{\bqa}{\begin{eqnarray}}
\newcommand{\eqa}{\end{eqnarray}}
\newcommand{\bea}{\begin{eqnarray}}
\newcommand{\eea}{\end{eqnarray}}

\newcommand{\N}{\mathds{N}}

\newtheorem{definition}{Definition}

\newtheorem{proposition}{Proposition}

\newcommand{\cF}{{\cal F}}

\newcommand{\cG}{{\cal G}}
\newcommand{\cB}{{\cal B}}
\newcommand{\cJ}{{\cal J}}

\newcommand{\cV}{{\cal V}}
\newcommand{\cE}{{\cal E}}

\DeclareMathOperator{\Tr}{Tr}

\DeclareMathOperator{\OO}{O}

\begin{document}

\title{\Large \bf Random tensor models in the large N limit:\\
\large Uncoloring the colored tensor models}

\author{{\bf Valentin Bonzom}}\email{vbonzom@perimeterinstitute.ca}
\author{{\bf Razvan Gurau}}\email{rgurau@perimeterinstitute.ca}
\affiliation{Perimeter Institute for Theoretical Physics, 31 Caroline St. N, ON N2L 2Y5, Waterloo, Canada}
\author{{\bf Vincent Rivasseau}}  \email{vincent.rivasseau@gmail.com}
\affiliation{Perimeter Institute for Theoretical Physics, 31 Caroline St. N, ON N2L 2Y5, Waterloo, Canada}
\affiliation{Laboratoire de Physique Th\'eorique, CNRS UMR 8627, Universit\'e Paris XI, 91405 Orsay Cedex, France}

\date{\small\today}

\begin{abstract}
\noindent Tensor models generalize random matrix models in yielding a theory of dynamical triangulations in arbitrary dimensions.
Colored tensor models have been shown to admit a $1/N$ expansion and a continuum limit accessible analytically.
In this paper we prove that these results extend to the most general tensor model for a single generic, i.e. non-symmetric, complex tensor. Colors appear in this setting as a canonical book-keeping device and not as a fundamental feature.
In the large $N$ limit, we exhibit a set of Virasoro constraints satisfied by the free energy and an infinite family of multicritical behaviors with entropy exponents $\gamma_m=1-1/m$.
\end{abstract}

\medskip

\noindent  Pacs numbers: 02.10.Ox, 04.60.Gw, 05.40-a

\maketitle

\section{Introduction}

Matrix models are probability measures for random matrices $M$ of size $N$. In physics language, they come with a matrix action $S(M)$. They can be divided in two broad categories. In the first category, that of  {\emph{invariant matrix models}} \cite{Di Francesco:1993nw},
the full action has an expansion in terms of traces of powers of $M$ (for Hermitian, or $\Tr (MM^\dagger)^n$ for general
$M$) which ensures invariance under $U(N)$ transformations.
The archetypes for this category are the $\lambda \Tr\,M^3$ or $\lambda \Tr\,M^4$ models, whose actions are
$S=\Tr M^2 + \lambda \Tr M^3$ or $S=\Tr M^2 + \lambda \Tr M^4$. The perturbative expansion of such models
involves ribbon graphs dual to triangulated, or quadrangulated, Riemann surfaces. Hence
(forgetting for a brief moment the constructive issues) these models are
statistical models of random {\emph{discretized}} Riemann surfaces. In the {\emph{large} $N$ limit, planar surfaces dominate
and furthermore undergo at some finite coupling a transition to {\emph{continuous}} surfaces \cite{Kazakov:1985ds,mm}, known as the large volume, or continuum limit.
Hence they provided until recently the only known example of analytically controlled geometrogenesis\footnote{This term has appeared for the first time in \cite{Konopka:2008hp} which is however quite different from our approach.}, i.e. the emergence of continuous geometries from discrete models, although restricted to two dimensions.
Moreover invariant single or multi-matrix models can also probe the critical behavior of two-dimensional statistical models on random geometries
\cite{Kazakov:1986hu, Boulatov:1986sb, Brezin:1989db,Kazakovmulticrit}\footnote{This critical behavior on random geometry is related to the one on fixed geometry through
the KPZ correspondence \cite{Knizhnik:1988ak, david2, DK, Dup}.}.

The second category of matrix models is that of {\emph{matrix field theories}}, in which the interaction is invariant, but the quadratic part
of the action is not. Since invariant $\Tr M^n$ interactions are the matrix analogs
of {\emph{local interactions}} $\int \phi^n (x)$, matrix field  theories
are the analogs of ordinary quantum field theories,  in which interactions are local but the propagator
(inverse of the Laplacian or Dirac operator) is not. From this point of view
invariant matrix models should be considered as {\emph{ultralocal}} matrix field theories.
Non-local propagators in field theory give birth to renormalization, hence to a flow
of the couplings.  Just as $\phi^4_4$ is the archetype for ordinary renormalization,
the archetype of matrix field theories is the
Grosse-Wulkenhaar model in four dimensions\footnote{The Grosse-Wulkenhaar model is a $\phi_4^{*4}$ model on the non-commutative Moyal space
with a harmonic potential. It does not suffer from the UV/IR mixing and becomes a matrix field theory
in the Moyal matrix base.}, or $GW_4$ \cite{Grosse:2004yu}.
This $GW_4$ model improves on the ordinary $\phi^4_4$ model since it is asymptotically safe  \cite{Disertori:2006nq}, hence free of the old Landau ghost problem.

Returning now to the important constructive question, let us recall that the constructive
analysis of stable invariant matrix models is compatible with their 1/N expansion. Borel summability has been proved
to hold uniformly in $N$ in the quartic case \cite{Rivasseau:2007fr}. For higher degree stable interactions a straightforward generalization of the techniques
of  \cite{Rivasseau:2010ke} should also lead to uniform Borel-LeRoy summability of the appropriate order. The constructive analysis of matrix
field theories is under
way \cite{Wang:2011hf} and expected to lead to a full construction of the $GW_4$ model in the near future.

All these nice properties of matrix models stem from their $1/N$ expansion
\cite{'tHooft:1973jz}, which states that planar graphs (dual to the sphere) govern their large $N$
limit\footnote{Through double scaling limits
one can even to some extent treat the sum over sub-leading terms in the $1/N$ expansion \cite{double,double1,double2}.}.
Planar graphs proliferate only exponentially in their number of vertices and
can be counted precisely through algebraic equations \cite{Brezin:1977sv}, as they are related to trees \cite{Tut,Sch,BMS}. This key feature
underlies all the statistical mechanics applications of the invariant models.
Renormalizability and asymptotic safety in the $GW_4$ model also rely entirely
on the dynamical analog of the $1/N$ expansion \cite{Grosse:2004yu, Disertori:2006nq, Gurau:2005gd, Rivasseau:2005bh}. Indeed in
such matrix field theories only planar graphs with a single external face look like matrix invariant terms at high energy, and
they are also the only ones to require renormalization.

Random matrices generalize in higher dimensions to random tensors \cite{ambj3dqg,sasa1,mmgravity} (and \cite{sasaa,sasab,sasac} for more recent developments),
whose perturbative expansion performs
a sum over random higher dimensional triangulations. But until recently all nice aspects of matrix models listed above could not be generalized
to tensors, as their $1/N$ expansion was missing. The situation has changed with the discovery of
{\it colored} \cite{color,PolyColor,lost} rank $D\ge 3$ random tensor models\footnote{In $D=2$ colors do not play the key role they play in three and more dimensions.
This is because there is a natural composition rule on rank-2 tensors, namely matrix multiplication, and a single trace invariant at order $n$, namely $\Tr M^n$.
In $D \ge 3$ there is no longer any multiplication law and there are many different invariants at order $n$. The
colors become essential as a canonical device to keep track of their combinatorics. Colored matrix models can of course still been defined
and have been studied in \cite{difrancesco-rect}.}.
These models require $D+1$ different
pairs of conjugate tensors $T^i, \bar T^i$, equipped with a particular
invariant canonical action of the type $\sum_i T^i\bar T^i + \lambda \prod_{i=0}^D T^i + \bar \lambda \prod_{i=0}^D \bar T^i$.
Their perturbation theory supports a $1/N$ expansion \cite{Gur3,GurRiv,Gur4}, indexed by the {\it degree},
a positive integer which plays in higher dimensions the
role of the genus but is not a topological invariant. Leading order graphs triangulate
the $D$-dimensional sphere in any dimension  \cite{Gur3,GurRiv}. These graphs, baptized {\it melonic} \cite{Bonzom:2011zz},
again proliferate only exponentially, as they map to colored $(D+1)$-ary trees \cite{Bonzom:2011zz,Gurau:2011tj}.
Like matrix models, these tensor models reach a continuum limit  when the coupling constant approaches its critical value.
The corresponding entropy exponent is $\gamma_{\rm melons} = 1/2$ in any dimensions \cite{Bonzom:2011zz}. It is the analog of the string susceptibility exponent $\gamma_{\rm string} = -1/2$ of the
invariant matrix models for the universality class of pure 2d quantum gravity,

Colored random tensors \cite{coloredreview} therefore gave the first theory of random geometries
in three and more dimensions with analytically tractable geometrogenesis and the subject is rapidly
expanding \cite{sefu2,Baratin:2011tg,Geloun:2011cy,BS3,Bonzom:2012sz,Ryan:2011qm,Carrozza:2011jn}.
Coupling of statistical mechanical systems to these random geometries in arbitrary dimension has been done in \cite{Bonzom:2012sz, IsingD,EDT},
and results at all orders in $1/N$ have been established for some restricted models \cite{doubletens} (see also
\cite{rey} for some related developments).

Obvious questions then arise. Do $1/N$ expansions also hold for uncolored models, i.e. with a single tensor?
How can one build tensor models with interactions of arbitrary degree that still admit a $1/N$ expansion?
What are the tensor analogs of matrix field theories?

A first important step towards answering the first two questions was taken in \cite{Gurau:2011tj}.
It was shown that integrating out all colored tensors
but one in the initial colored model leads to an effective action for a single {\emph{uncolored}} tensor which is a sum of
effective  invariant interactions whose internal structure can be unfolded in terms of colored graphs.

In the present paper we return to these questions in greater detail.
Like for matrices, we can distinguish invariant tensor models and tensor field theories.
Invariant tensor models are those considered in this paper. They correspond to tensors with both
quadratic part and interactions invariant under the external tensor product $\otimes^{D} U(N)$. We consider the most general invariant models for a \emph{random, complex tensor}. It is important that this tensor is generic, that is without any symmetrization or antisymmetrization of its indices.
Labeling these indices then provides exactly the same combinatorial tool that colors provide in the colored models.
It allows us to
\begin{itemize}
\item define their $1/N$ expansion, again organized according to the degree of the graphs,
\item prove it is dominated by melonic, colored graphs of spherical topology,
\item derive the continuum limit, whose entropy exponent is generically $\gamma_{\rm melons}=1/2$, thus proving the universality of this continuum phase\footnote{It is analogous to the universality class of pure 2d quantum gravity which is obtained for most values of the coupling constants in one-matrix models \cite{Di Francesco:1993nw, Kazakovmulticrit}.},
\item extract a set of Virasoro constraints which hold in the large $N$ limit,
\item find multicritical points, with entropy exponents $\gamma_m = 1 - \frac{1}{m}$ (for $m\geq2$ integers), which are the same as the ones of multicritical branched polymers \cite{BP-ambjorn,doubletens, Bonzom:2012sz}. This is the generalization of \cite{Kazakovmulticrit} to tensors.
\end{itemize}
These are the main results of this paper, and they are direct consequences of the \emph{universality} of tensor invariant measures first derived in \cite{Gurau:2011kk}. We stress that universality in this context \emph{only} means that in the large $N$ limit the tensors are distributed on a Gaussian. However, the Gaussian itself (i.e. its covariance) is \emph{not} universal, but depends on the coupling constants of the model. Indeed, when the large $N$ covariance becomes critical, the continuum limit is reached. Further, tuning the couplings appropriately, multicritical behaviors are observed, just like in invariant matrix models \cite{Kazakovmulticrit}.

We expect the constructive analysis of stable and symmetric invariant models not to pose any difficulty, as the necessary techniques have been in fact already developed for the quartic case \cite{Magnen:2009at} in the slightly different context of group field theory \cite{Oriti:2011jm}.

Tensor field theories are the analogs of matrix field theories. They have tensor invariant interactions but a Laplacian-based propagator.
Such a propagator again allows a renormalization group analysis. We do not consider this second category of models further in this paper,
except to recall that {\emph{uncolored}}
renormalizable models of this type have been found for rank 3 and rank 4 tensors
\cite{BenGeloun:2011rc,BenGeloun:2012pu}.
Again the renormalization in such models is entirely based on a dynamical version of the $1/N$ expansion.
One should explore their flows, phase transitions, critical exponents, gauge invariant extensions and constructive properties,
as they seem a promising approach to the quantization of gravity in more than two dimensions \cite{Rivasseau:2011hm}.

We follow the standard presentation of invariant matrix models in the large $N$ limit, like in the well-known review \cite{Di Francesco:1993nw}, to emphasize the new status of the field.
In section \ref{sec:model}, we define the generic models. In section \ref{sec:largeN} we consider their $1/N$ expansion, which is dominated by the melonic graphs and establish their continuum limit. In section \ref{sec:multic} we analyze the infinite family of multicritical points for these models.

\section{The $1/N$ Expansion of Invariant Tensor Models}\label{sec:model}

\subsection{Tensor invariants and action}

The models we consider are based on complex tensors which have \emph{no} symmetry between their indices. In order to write the most general action,
one must first understand the invariants built from such tensors. It turns out that the analysis of these invariants automatically leads to a representation
 in colored graphs.

Let $H_1,\dotsc,H_D$ be complex vector spaces of dimensions $N_1,\dotsc,N_D$. A rank $D$ covariant tensor $T_{n_1\dotsc n_D}$ can be seen as a collection of $\prod_{i=1}^{D} N_i$ complex numbers supplemented with the requirement of covariance under base change. We consider tensors $T$
transforming under the {\it external} tensor product of fundamental representations of the unitary group $\otimes_{i=1}^D U(N_i)$,
that is each $U(N_i)$ acts independently on its corresponding $H_i$.
The complex conjugate tensor $\bar T_{  n_1 \dotsc n_D }$ is then a rank $D$ contravariant tensor. They transform as
\bea
 T'_{a_1\dotsc a_D} = \sum_{n_1,\dotsc,n_D} U_{a_1n_1}\dotsm V_{a_Dn_D}\ T_{n_1\dotsc n_D}  \; ,\qquad
 \bar T'_{  a_1\dots  a_D} = \sum_{n_1,\dotsc,n_D} \bar U_{  a_D n_D}\dotsm \bar V_{a_1 n_1}\ \bar T_{n_1\dots n_D}  \; .
\eea
From now on we will always denote the indices of the complex conjugated tensor with a bar.
We will sometimes denote the $D$-uple of integers $(n_1, \dotsc, n_D)$ by $\vec n$ and assume (unless otherwise specified) $D\ge 3$.
We restrict to $H_i =H$, $N_i =N$, for all $i$.

Among the polynomial quantities one can build out of $T$ and $\bar T$ we will deal in the sequel exclusively with
\emph{trace invariants}. The trace invariants are built by contracting two by two
covariant with contravariant indices in a polynomial in the tensor entries.
We write trace invariants formally like
\bea
\Tr (T,\bar T) = \sum \prod \delta_{n_1,\bar n_1} \;  T_{n_1\dotsc} \dots \bar T_{\bar n_1 \dots } \; ,
\eea
where \emph{all} indices are saturated. Note that a trace invariant has necessarily the same number of
$T$ and $\bar T$.

Trace invariants can be labeled by graphs with distinguished vertices. To draw the graph associated to a trace invariant we
represent every $T$ by a white vertex $v$ and every $\bar T$ by a black vertex $\bar v$. We promote the position of an index
to a color: $n_1$ has color $1$, $n_2$ has color $2$ and so on. The contraction of two indices $n_i$ and $\bar n_i$ of tensors
is represented by a line $l^i = (v,\bar v)$ connecting the corresponding two vertices. Lines inherit the color of the index, and
always connect a black and a white vertex. Any trace invariant is then represented by a $D$-colored graph.
\begin{definition}
A {\bf closed $D$-colored graph}, or {\bf $D$-bubble}, is a graph $\cB = (\cV,\cE)$ with vertex set $\cV$ and line set $\cE$
such that
\begin{itemize}
\item  $\cV$ is bipartite, i.e. there exists a partition of the vertex set $\cV  = A \cup \bar A$, such that for any
element $l\in\cE$, then $ l = \{v,\bar v\}$ with $v\in A$ and $\bar v\in\bar A$. Their cardinalities
satisfy $|\cV| = 2|A| = 2|\bar A|$.
\item  The line set is partitioned into $D$ subsets $\cE = \bigcup_{i  =1}^{D} \cE^i$, where $\cE^i$ is the subset
of lines with color $i$, with $|\cE^i|=|A|$.
\item  It is $D$-regular (all vertices are $D$-valent) with all lines
incident  to a given vertex having distinct colors.
\end{itemize}
\end{definition}

Some examples of trace invariants for rank 3 tensors are represented in figure \ref{fig:tensobs}.
\begin{figure}[t]
\begin{center}
 \includegraphics[width=8cm]{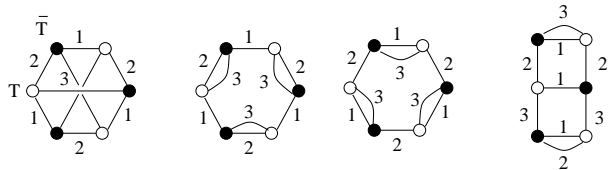}
\caption{Graphical representation of trace invariants.}
\label{fig:tensobs}
\end{center}
\end{figure}
The trace invariant associated to the graph $\cB$ writes as
\bea
\Tr_{\cB}(T,\bar T ) = \sum_{\{\vec n^v,\bar{\vec{n}}^v\}_{v,\bar v \in \cV}}  \delta^{\cB}_{\{\vec{n}^v, \bar{\vec{n}}^{\bar v}\}}  \;
\prod_{v,\bar v \in \cB} T_{\vec n^v} \bar T_{\bar {\vec n }^{\bar v} }
\; ,\quad \text{with} \qquad \delta^{\cB}_{\{\vec{n}^v, \bar{\vec{n}}^{\bar v}\}} = \prod_{i=1}^D \prod_{l^i = (v,\bar v)\in \cB} \delta_{n_i^v \bar n_i^{\bar v}} \; .
\eea
where $l^i$ runs over all the lines of color $i$ of $\cB$. We call the $\delta^{\cB}_{\{\vec{n}^v, \bar{\vec{n}}^{\bar v}\}}$
(the product of delta functions encoding the index contractions of the observable associated to the graph $\cB$)
the \emph{trace invariant operator with associated graph} $\cB$ \cite{Gurau:2011tj}. Trace invariant operators factor over the connected
components of the graph. From now on we will always consider connected invariants, hence invariants associated to connected graphs in the above
representation. We denote $\Gamma_{2k}^{ ( D ) }$ the set of $D$-colored, connected graphs with $2k$ distinguished vertices and
$\Gamma^{(D)}$ the set of all graphs with $D$ colors.

Of particular importance in the sequel are the subgraphs with two colors of a $D$-colored graph, called \emph{faces}. We denote them $\cF$.
For instance the graphs with $3$ colors posses three type of faces, given by the subgraphs with lines of colors $12$, $13$ and $23$.
As every line belongs to exactly two faces (the lines of color $1$ belong to a face $12$ and a face $13$, etc.), the graphs with
three colors can be represented as ribbon graphs.

To every graph $\cB$ with $D$ colors we can associate a non-negative integer,
its \emph{degree} $\omega(\cB)$ \cite{GurRiv,Gur4,coloredreview}. We recall its definition and properties in the appendix \ref{app}. The main feature of the
degree is that it provides a counting of the number of faces of a graph, thus for a graph with $D$ colors and $2p$ vertices the total
number of faces computes
\bea \label{eq:facesmese}
  |\cF| = \frac{(D-1)(D-2)}{2} p + (D-1) - \frac{2}{(D-2)!} \omega(\cB) \; .
\eea
Taking into account that graphs with $3$ colors are ribbon graphs, it is easy to see that in this case the degree reduces
to the genus. In higher dimensions the degree provides a generalization of the genus. It is {\it not} a topological invariant,
but it combines topological and combinatorial information about the graph.

Going back to invariants one can build out of a complex tensor, we note that there exists a unique $D$-colored graph with
two vertices, namely the graph in which all the lines connect the two vertices. We call it the $D$-dipole (denoted $\cB_1$)
and its associated invariant is
\bea\label{eq:gaussian}
 \Tr_{\cB_1} (T , \bar T ) = \sum_{\vec n,\bar{\vec{n}}}\, T_{\vec n}\, \bar T_{\bar {\vec n} }\ \Bigl[\prod_{i=1}^D \delta_{n_i \bar n_i}\Bigr] \; .
\eea
The most general invariant action for a non-symmetric tensor is therefore
\bea \label{eq:actiongen}
 S(T,\bar T) = t_1\, \Tr_{\cB_1} (T , \bar T ) + \sum_{k=2}^{\infty} \sum_{\cB\in \Gamma_{2k}^{ ( D )} } t_{\cB}\,
N^{-\frac{2}{(D-2)!} \omega(\cB)} \,
\Tr_{\cB}(T,\bar T)\;,
\eea
where $(t_\cB)$ is the set of coupling constants associated to $D$-bubbles and we singled out the quadratic part corresponding to $\cB_1$.
In equation \eqref{eq:actiongen} we have added a scaling in $N$ for every trace invariant, proportional to its degree. As the degree is non-negative
this scaling is a suppression of some invariants. We have included it because it simplifies some equations in the following, but we emphasize that this scaling is {\it not} required: as the reader can check all the
results we present below generalize (albeit with some effort) also in its absence. Due to symmetry under relabeling of the black and white vertices, some couplings in \eqref{eq:actiongen} are redundant. It is however
more convenient to assign a distinct coupling constant
to each graph with labeled vertices, and remember this redundancy only at the end.

We will deal in this paper with the most general single-tensor model of
rank $D$ defined by the partition function
\be
Z(t_\cB) = \exp\bigl(-F(t_\cB)\bigr) = \int d\bar T dT\ \exp\ \Bigl(-N^{D-1} S (T,\bar  T)\Bigr) \;.
\ee

\subsection{Graph amplitudes}

The invariant observables are the trace invariants represented by $D$-colored graphs. The Feynman graphs
contributing to the expectation of an observable are obtained by Taylor expanding with respect to $t_{\cB}$ and evaluating
the Gaussian integral in terms of Wick contractions. A moment of reflection reveals that the Feynman graphs
are made of {\it effective vertices} $ \Tr_{\cB}(T,\bar T) $ (that is graphs $\cB$ with colors $1,\dotsc, D$)
connected by effective \emph{propagators} (Wick contractions, pairings of $T$'s and $\bar T$'s).
A Wick contraction of two tensor entries $T_{a_1\dotsc a_D}$ and $\bar T_{\bar p_1 \dotsc \bar p_D}$
with the quadratic part \eqref{eq:gaussian} consists in replacing them by
$ \frac{ 1}{N^{D-1} t_1} \prod_{i=1}^D \delta_{a_i \bar p_i} $.
The Wick contractions will be represented as dashed lines labeled by the fictitious color $0$.
Thus every dashed line of color $0$ in a Feynman graph identifies \emph{all} the indices of the two vertices
(one white corresponding to $T$ and one black corresponding to $\bar T$) it connects.
An example of a Feynman graph is presented in figure \ref{fig:tensobsgraph}.

\begin{figure}[t]
\begin{center}
 \includegraphics[width=3cm]{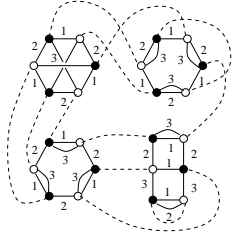}
\caption{A Feynman graph.}
\label{fig:tensobsgraph}
\end{center}
\end{figure}

The Feynman graphs are therefore $(D+1)$-colored graphs $\cG$. We reserve the notation $\cB$ for the $D$-colored graphs,
and $\cG$ for the $(D+1)$-colored graphs.
A graph $\cG$ has two kinds of faces: those with colors
$i,j=1,\dotsc,D$, denoted $\cF_{ij}$ (which belong also to some D-bubble $\cB$) and those with colors $0,i$, for $i=1,\dotsc,D$, denoted $\cF_{0i}$,
which involve the lines of color 0 in $\cG$.

The free energy has an expansion in closed, connected $(D+1)$-colored graphs,
\be
F(t_{\cB}) = \sum_{\cG\in \Gamma^{(D+1)}} \frac{ (-1)^{|\rho|}}{s(\cG)}\ A(\cG) \;,
\ee
where $s(\cG)$ is a symmetry factor and $|\rho|$ is the number of effective vertices i.e. $D$-bubbles (subgraphs
with colors $1,\dotsc, D$). We denote these $D$-bubbles $\cB_{(\rho)}$, with $\rho=1,\dotsc, |\rho|$.
The amplitude of a graph is
\be
A(\cG) = \prod_{\rho} t_{\cB_{(\rho)}} \; \sum_{\{\vec{n}^v,\bar{\vec{n}}^{\bar v}\}}\, \Bigl[
\prod_{\rho}  N^{D-1 -\frac{2}{(D-2)!} \omega(\cB_{(\rho)} )   } \delta^{\cB_{(\rho)}}_{\{\vec{n}^v,\bar{\vec{n}}^{\bar v}\}}\Bigr]
\Bigl[\prod_{l^0=(v,\bar v) \in \cE^0} \frac{1}{t_1 N^{D-1}} \prod_{i} \delta_{n_i^v,\bar n_{i}^{\bar v}} \Bigr]\;.
\ee
An index $n_i$ is identified along the lines of color $i$ in $\cB_{(\rho)}$ and along the dashed lines of color $0$. We thus obtain a free
sum per face of colors $0i$, so that
\be
A(\cG) =  \frac{\prod_{\rho} t_{\cB_{(\rho)}}}{t_1^{ |l^0| }}
   \ N^{(D-1)|\rho| - \frac{2}{(D-2)!} \sum_{\rho} \omega(\cB_{(\rho)} )    - (D-1) |l^0| + \sum_i |\cF_{0i}| } \;.
\ee
Noting that $\sum_{i}|\cF_{0i}|= |\cF|- \sum_{\rho}|\cF_{(\rho)}|$, where $|\cF|$ denotes the
total number of faces of $\cG$ and $|\cF_{(\rho)}|$ the number of faces of the $D$-bubble $\cB_{(\rho)}$,
using \eqref{eq:facesmese} for each $\cB_{(\rho)}$ and for $\cG$ (taking into account that $\cG$ has $D+1$ colors)
and noting that $|l^0|=p$, with $p$ the half-number of vertices of $\cG$, the amplitude of $\cG$ computes
\be\label{eq:amplifini}
A(\cG) = \frac{\prod_{\rho} t_{\cB_{(\rho)}}}{t_1^{ p }} \ N^{D -\frac{2}{(D-1)!} \omega(\cG)  }\;,
\ee
with $\omega(\cG)$ the degree of the graph $\cG$. The $1/N$ expansion of the free energy writes
\be
F(t_{\cB}) = N^D \; \sum_{\cG \in \Gamma^{(D+1)} }
\frac{ (-1)^{|\rho|}}{s(\cG)} \frac{\prod_{\rho} t_{\cB_{(\rho)}}}{ t_1^{p} } \; N^{ -\frac{2}{(D-1)!} \omega(\cG)}
\; .
\ee
The leading scaling with $N$ of the free energy is $F(t_{\cB})\sim N^D$.
In the rest of this paper we focus on the leading order free energy $f_0(t_{\cB}) = \lim_{N\to \infty} N^{-D} F(t_{\cB}) $.
Expectation values of bubble observables have similar expansions.
If $\cB$ is a $D$-colored graph, the connected expectation value
\begin{align}\label{eq:observables}
\frac{1}{N} \frac{ \Big{\langle}  \Tr_{\cB}(T,\bar T) \Big{\rangle} }{Z}
= \frac1{N^D}\ N^{\frac{2}{(D-2)!} \omega(\cB) }\; \frac{\partial F}{\partial t_\cB}
= \sum_{\cG \in \Gamma^{(D+1)}, \cG \supset \cB } \frac{(-1)^{|\rho|}}{s(\cG)}\
\frac{\prod_{\rho} t_{\cB_{(\rho)}}}{ t_1^{p} } \; N^{ -\frac{2}{(D-1)!} \omega(\cG) + \frac{2}{(D-2)!} \omega(\cB)} \; ,
\end{align}
has an expansion in connected $(D+1)$-colored vacuum graphs $\cG$ having $\cB$ as a (marked) subgraph, denoted $\cG \supset \cB$.
The scaling in $N$ in \eqref{eq:observables} of a graph $\cG$ rewrites
\bea
  N^{-\frac{2}{D!} \omega(\cG)} N^{-\frac{2}{D(D-2)!} \bigl( \omega(\cG) - D\, \omega(\cB) \bigr) } \; .
\eea
Using (a weaker version of) proposition \ref{deg>deg} in appendix \ref{app}, $\omega(\cG) \ge D\, \omega(\cB)$ and the inequality is
saturated for $\omega(\cG)=0$. It follows that in the large $N$ limit only graphs $\cG\supset \cB$ of degree zero contribute to
the expectation.

\subsection{Topology from bubbles}

To simplify the discussion, in this section we will restrict to the case $D=3$. The original idea of tensor models \cite{mmgravity,sasa1,ambj3dqg}
was to generate triangulations of $3$-dimensional spaces. The basic building block in the original proposals was an interaction term
which combinatorially describes a tetrahedron (a $3$-simplex) also used in group field theories \cite{Oriti:2011jm}
\be
V_{\text{tetrahedron}} = \sum_{a,b,c,d,e,f} T_{abc} T_{cde} T_{ebf} T_{fda} \; .
\ee
This term is not $\otimes^3 U(N)$ invariant. The most one can say about it is that it is invariant under
a simultaneous $\OO(N)$ orthogonal transformation of all its indices.

The situation is already improved in colored tensor models \cite{color} where the indices are distinguished and one can implement a $\otimes^3 U(N)$
invariance. As the pattern of contraction of a tetrahedron is not a trace invariant one can raise the question of the topological interpretation of
the trace invariant observables and their relation to triangulations.

The situation is actually like in one-matrix models with generic interactions. A $\Tr(M^k)$-vertex is seen (by duality) as a polygon with $k$ sides.
A closed graph is then a gluing of such polygons. Obviously one can divide each polygon into triangles (by adding a vertex in the middle of
the polygon, i.e. by taking the topological cone over its boundary), so that the graph encodes a triangulation.
Here, a similar interpretation holds. The $(3+1)$-colored graphs are known to describe topological $3$-dimensional pseudo-manifolds \cite{color}.
The black and white vertices of the graph correspond to tetrahedra ($3$-simplices). The triangles ($2$-simplices) bounding a tetrahedron
are represented by the half-lines touching the vertex, hence are colored $0,1,2,3$. The lower dimensional simplices are colored by the
colors of the triangles sharing them. Thus the edges are labeled by pairs of colors (the edge $12$ is common to the
triangles $1$ and $2$), and the points (vertices of the tetrahedra, to be distinguished from the vertices of the graph) are labeled by triples of colors (the point $123$ is the point common to the triangles $1$, $2$ and $3$ bounding a tetrahedron).

A line in the colored graph represents the unique gluing of two tetrahedra of opposite orientations along boundary triangles
which respects {\it all} the colorings: that is we glue triangles of the same color, say $2$, in such a way that the edge
$02$ (resp. $12$ and $32$) bounding a triangle is glued on the edge $02$ (resp. $12$ and $32$) bounding the second triangle, and similarly for points.
This construction yields the pseudo-manifold dual to a $(3+1)$-colored graph.

\begin{figure}[t]
\begin{center}
\subfigure[]{
\includegraphics[scale=0.16]{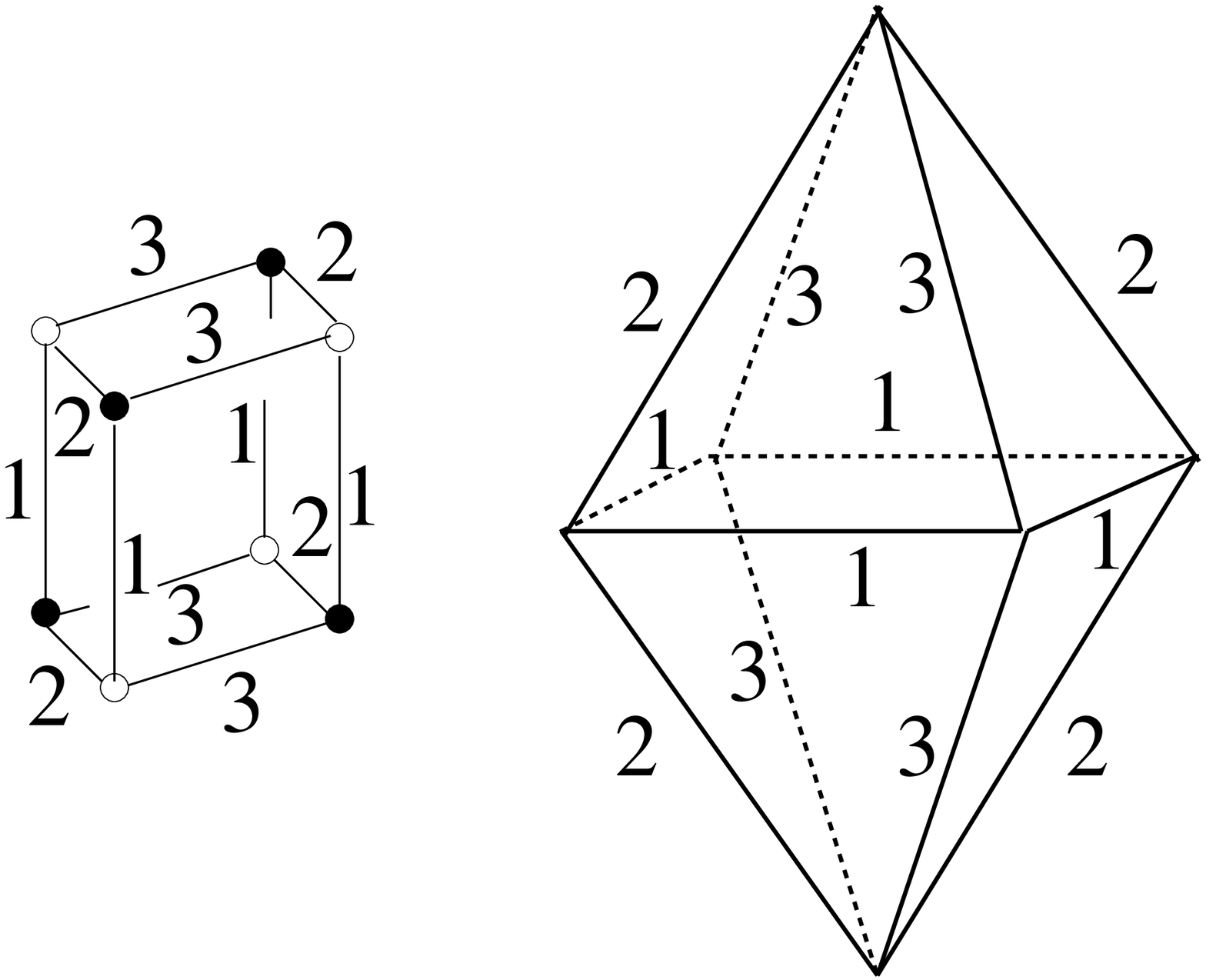}
  \label{fig:polytope1} }
  \hspace{1cm}
\subfigure[]{
\includegraphics[scale=0.16]{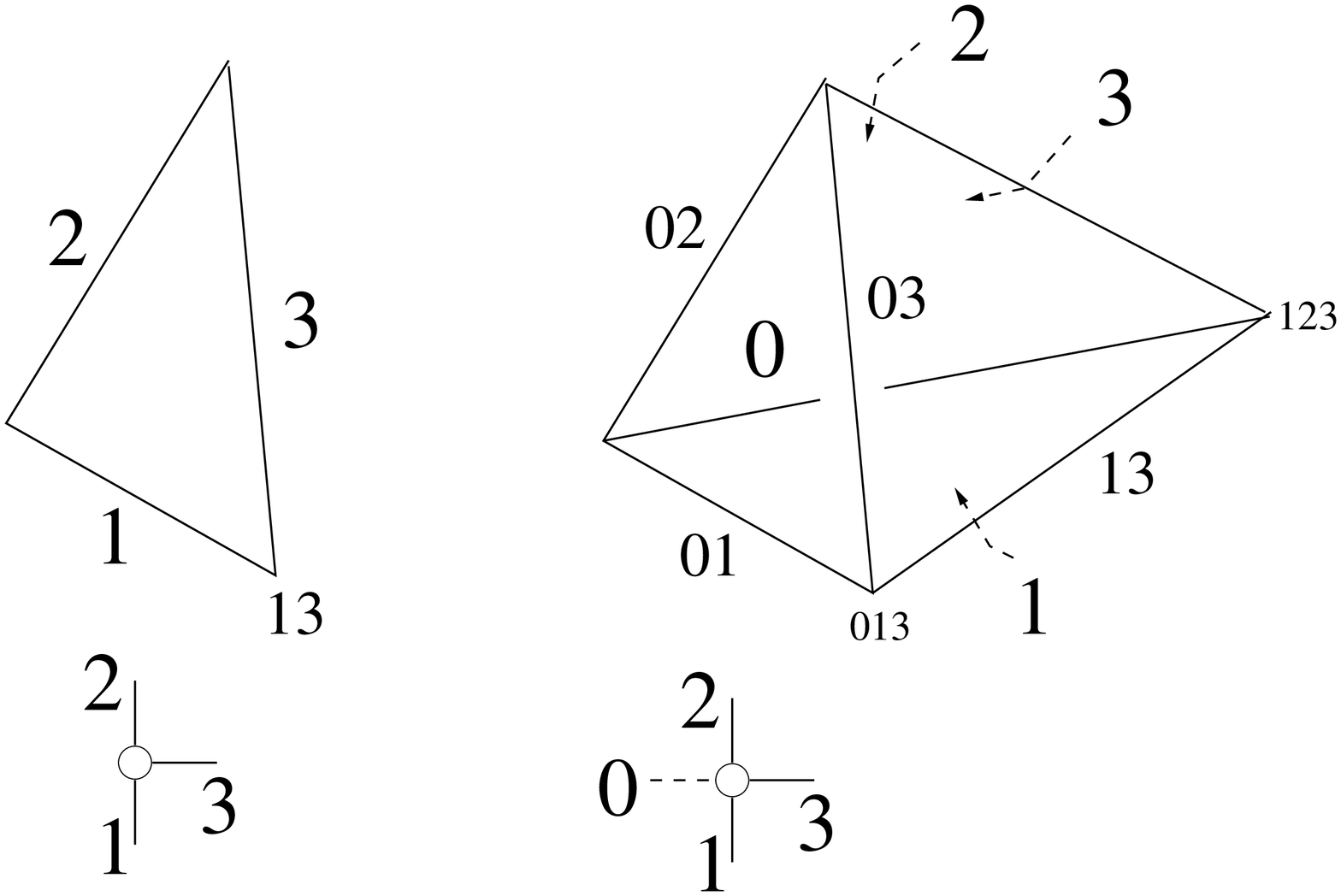}
  \label{fig:coning} }
  \hspace{1cm}
\subfigure[]{
\includegraphics[scale=0.16]{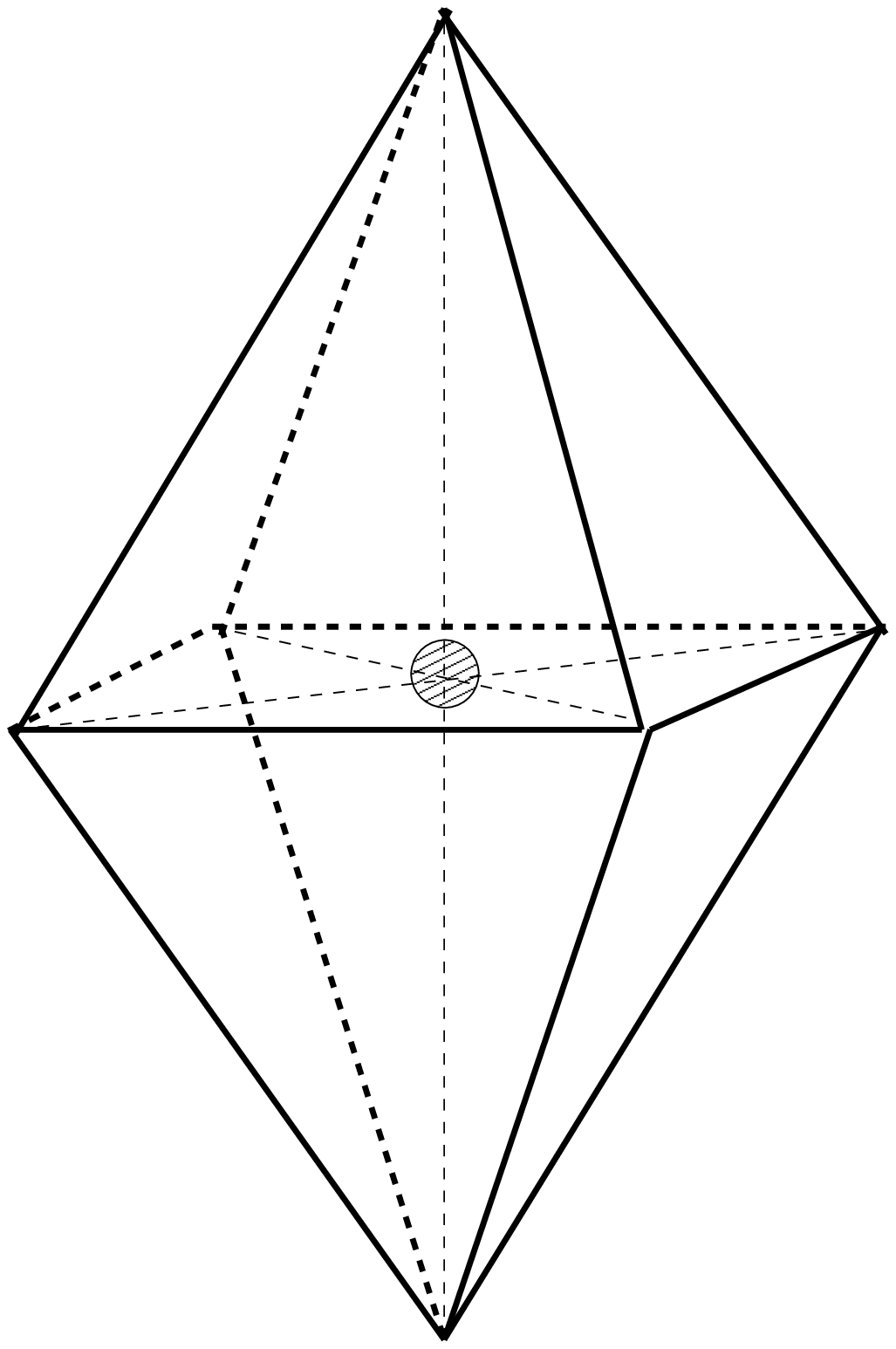}
  \label{fig:polytope} }
\hspace{1cm}
\subfigure[]{
\includegraphics[scale=0.4]{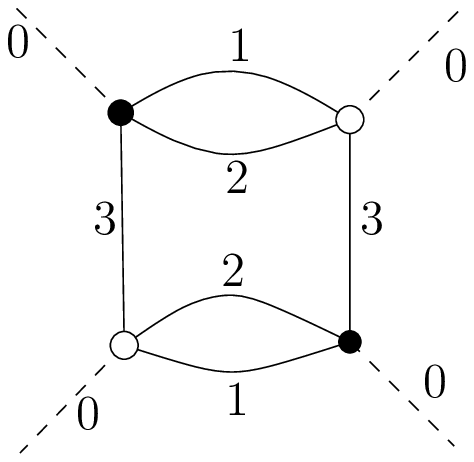}
\label{fig:uncolT4} }
\hspace{1cm}
\end{center}
\caption{Trace invariants and gluings of simplices in $D=3$.}
\end{figure}

Alternatively the same graph with $3+1$ colors can be seen as the gluing of the effective interactions, $\cB$ which are graphs
with $3$ colors, along the lines of color $0$. Following the above construction, each effective interaction by itself, being a graph with $3$ colors, represents a surface.
The (black and white) vertices are dual to triangles, and the edges bounding the triangles are colored $1$, $2$ and $3$.
The surface represented
by an interaction is the unique one obtained by gluing the triangles along their edges (as indicated by the graph with three colors) respecting
all the colorings (i.e. those of the edges and of the points).
In figure \ref{fig:polytope1} for instance we represented such a surface obtained by gluing eight triangles.

Adding the lines of color $0$ results in taking the topological cone over this pseudo-manifold, $C_M = (M\times [0,1])/ (M\times \{1\})$.
Let us first examine the effect of this coning on one triangle (represented in figure \ref{fig:coning}). The original triangle will
now be called a triangle of color $0$ (see figure \ref{fig:coning}). The original edges acquire the new color
$0$, hence they will be called $01$, $02$ and $03$ (see again \ref{fig:coning}), and similarly the original points ($13$ becomes $013$, etc.).
This coning adds extra triangles, edges and points. Every original edge gives by coning a new triangle. We color this triangle by the color of the edge,
hence the original edge of color $1$ gives rise by coning to the triangle of color $1$ (see again figure \ref{fig:coning}).
Note that the new triangle $1$ shares with the original triangle, $0$ the edge $01$. Every original point gives
by coning an edge, which inherits the colors of the original point (the edge $13$ is the cone over the original point $13$ and is shared by the triangles
$1$ and $3$). We also obtain a new point labeled $123$. When taking the cone over the surface defined by a connected graph
$\cB$ (with colors $1$, $2$ and $3$), we obtain new triangles (one for every edge of the surface), new edges (one for every point of the surface), and an unique new point $123$, the apex of the cone.

Thus, when seen as a subgraph of a $(3+1)$-colored graph $\cG$, $\cB$ represents a ``chunk'' of the $3$-dimensional space.
For the example of the graph with $3$ colors in figure \ref{fig:polytope1} adding the dashed lines of color $0$, we obtain
the gluing of $8$ tetrahedra drawn in figure \ref{fig:polytope}. This chunk has the topology of a ball
and is bounded by the $8$ triangles of color $0$ corresponding to the dashed half-lines.

Thus the trace invariant quartic interactions like
\be
\sum_{n_i,m_i} T_{n_1 n_2 n_3}\,\bar T_{n_1 n_2 m_3}\,T_{m_1 m_2 m_3}\,\bar T_{m_1 m_2 n_3}\; ,
\ee
represented in figure \ref{fig:uncolT4}, correspond to a gluing of four tetrahedra, with four external, boundary triangles of color $0$,
and not to a tetrahedron.
Note that a chunk can have a non-trivial topology, for instance it can be a cone over a torus.

One can employ an alternative {\it stranded graph} representation of the Feynman graphs, closer to the ribbon graph representation of matrix models.
This is presented in figure \ref{fig:stranded}. One replaces the black and white vertices of the effective interactions by stranded half-lines,
which are then connected by dashed lines having each three strands. In this representation the strands colored $1$, $2$ and $3$ have each an associated Kronecker $\delta$ which corresponds to the contraction of a tensor index between two tensors of the bubble observable. The dashed lines have three strands representing the three Kronecker
$\delta$ coming from a Wick contraction which propagate the tensor indices. The faces of colors $0i$ are easily identifiable.
Each stranded half-line corresponds to a triangle (the triangles of color $0$ in the colored graph representation).
The graph of the effective interaction encodes the pattern of gluing of the triangles into a surface (the boundary of a chunk), and
the dashed lines encode the gluing of chunks along boundary triangles. As this representation is redundant and somewhat cumbersome
we will not use it further.

Before concluding this section let us remark that the fact that the graphs are bipartite plays a secondary role, ensuring just the
orientability \cite{caravelli}.  What is crucial is that a colored
graph represents the unique gluing of simplices which respects {\em all} the labellings (including the induced ones over all the lower dimensional simplices). Dropping the bipartite requirement allows one to consider the $O(N)^3$ invariant presented in figure \ref{fig:fake}.
As it consists in a gluing of four triangles and any two triangles share exactly one edge one might be tempted to interpret it as a
gluing pattern of four triangles bounding a tetrahedron.
However, this interpretation is not the correct one. Indeed, on closer inspection, it turns out that the dual gluing consists in four triangles
glued first around a vertex (say $13$) and then glued along opposite edges of color $2$ (see the right hand side of figure \ref{fig:fake}). Thus,
respecting the rules of the colored gluings described above, this $O(N)^3$ invariant has the topology of the real projective plane
$\mathbb{R}P^2$.

\begin{figure}
  \subfigure[]{
\includegraphics[scale=0.3]{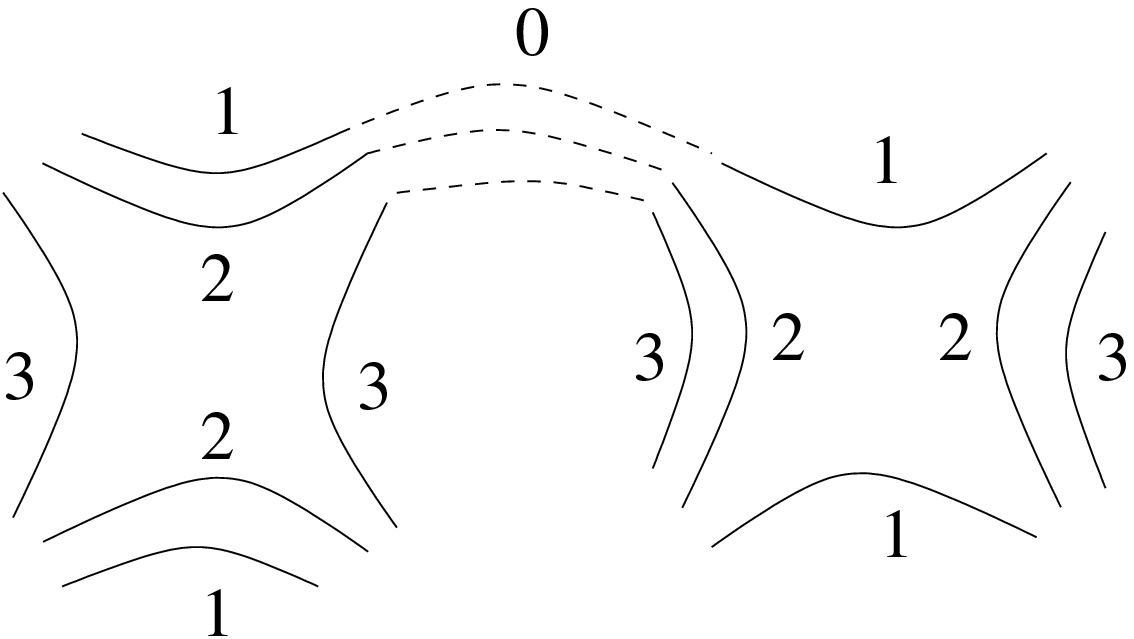}
\label{fig:stranded} }
\hspace{1cm}
\subfigure[]{
\includegraphics[scale=0.4]{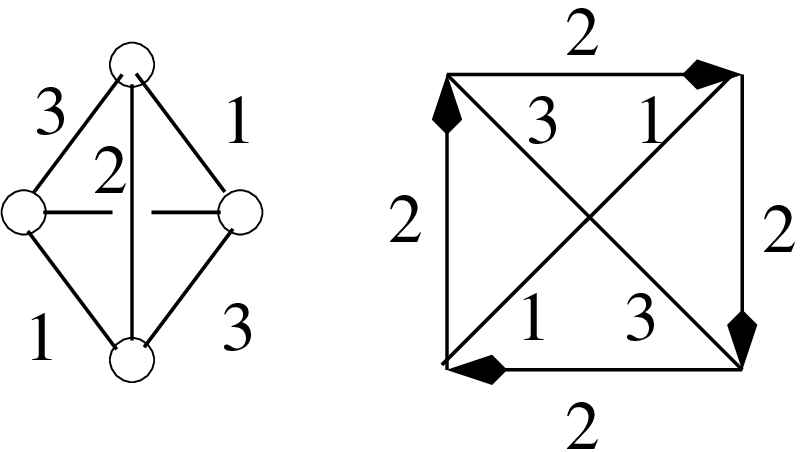}
\label{fig:fake} }
\caption{Stranded graph and a non-bipartite invariant.}
\end{figure}

\section{Large N limit}\label{sec:largeN}

\subsection{The melonic family and the large $N$ factorization}

In the large $N$ limit, only graphs with vanishing degree survive. For $(2+1)$-colored graphs the degree is the genus of the graph, hence the
graphs of degree $0$ are exactly the planar graphs and represent spheres. For $D\geq 3$, the $(D+1)$-colored graphs $\cG$
with $\omega(\cG)=0$ have been shown to also describe topological spheres in dimension $D$ \cite{Gur4,Bonzom:2011zz,coloredreview}.

\subsubsection{Combinatorial description of melons}

We explain in appendix \ref{app} why the $(D+1)$-colored graphs of degree zero, called \emph{melonic}, can be obtained by the insertion procedure detailed below. While the dominant graphs of our models are melonic, it is understood that not all melonic graphs are generated. However, this section is only concerned with the combinatorial properties of the melonic family, hence we temporarily allow ourselves to also use melonic graphs which do not appear in the Feynman expansion of our models. 

{\bf First order.}
The lowest order graph consists in two vertices connected by $D+1$ lines, as in the figure \ref{fig:order1}. We consider all lines
incident at the positive (black) vertex to be {\it active}, which means that higher order graphs will be obtained by insertions on them.

{\bf Second order.}
$D+1$ graphs contribute to the second order. They arise from inserting two vertices connected by $D$ lines
on any of the $D+1$ active lines of the first order graph. If the line on which we insert this decoration has color $i$,
the new lines will be colored by all colors except $i$. Say we insert this graph on the active line of color $1$, like in the
figure \ref{fig:order2} (hence the new lines have colors $2$, $3$ up to $D$). All lines incident at the new black vertex
are deemed active (the new lines of colors $2,\dotsc, D$ as well as the external line of color $1$), while the exterior line of color $1$ incident at the new white vertex (in bold in figure \ref{fig:order2}) is deemed inactive.

\begin{figure}[htb]
\subfigure[The first order melonic graph.]{\hspace{1cm}\includegraphics[width=2cm]{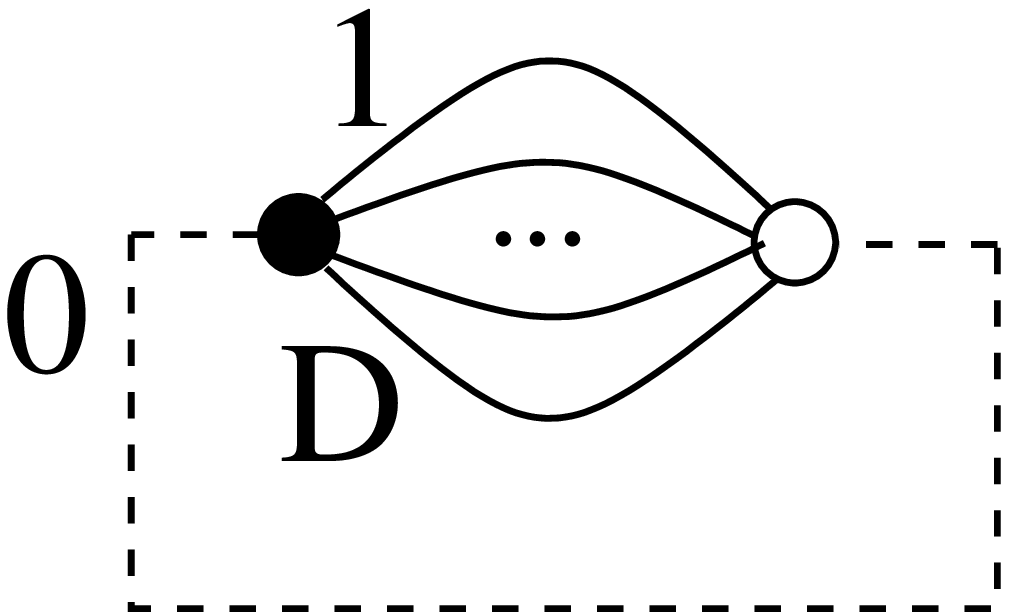} \label{fig:order1}\hspace{1cm}}\hspace{2cm}
\subfigure[A second order melonic graph.]{\hspace{1cm}\includegraphics[width=2cm]{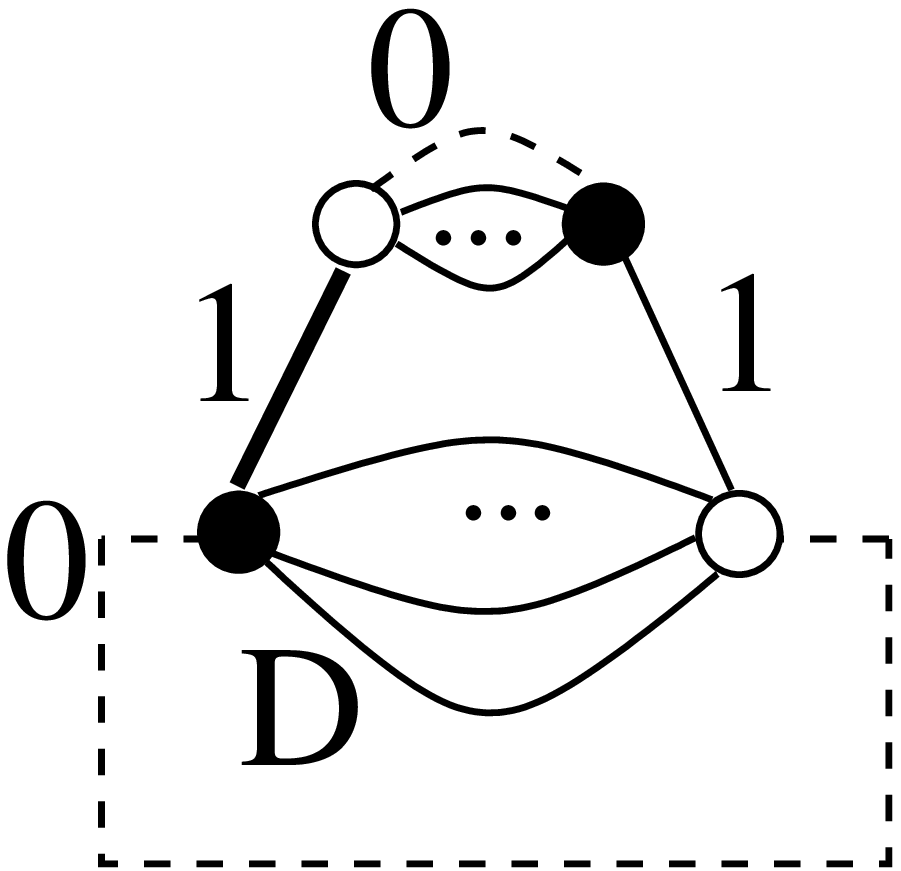} \label{fig:order2}\hspace{1cm}}
\caption{}
\end{figure}

{\bf Order $p+1$.}
We obtain the graphs at order $p+1$ by inserting two vertices connected by $D$ lines (with appropriate colors) on any of the active lines of a
graph at order $p$. Once again, with respect to the new vertices, all lines incident to the black vertex are deemed active, while the
exterior line incident to its white vertex is deemed inactive.

We are now going to show that the expectation values of melonic graphs are fully determined by the (dressed) covariance of the model, in a specific, factorized form.

\subsubsection{Large $N$ factorization}

In the large $N$ limit, only the bubble observables $\cB$ for which there exist $(D+1)$-colored graphs $\cG$ which are melonic
survive. The melonic graphs have some important properties, which put together lead to the large $N$ factorization of
expectations. 
\begin{itemize}
\item If a $(D+1)$-colored graph $\cG$ is melonic then all its subgraphs $\cB$ with colors $1,2,\dotsc, D$ are melonic
see figure \ref{fig:factorizationobs} and are therefore built following the same procedure.
\item In this procedure, $\cG$ is obtained by inserting pairs of vertices $v$ and $\bar v$ separated by $D$ lines, and a $D$-colored subgraph $\cB$ is obtained by performing the same insertions, but ignoring the color $0$.
\item Consider two vertices $v$ and $\bar v$ inserted at some step. At the time of the insertion, they are connected in $\cG$ by $D$ lines and some two-point graph (corresponding to the line on which they have been inserted). As all further insertions are made on the lines of $\cG$, the two half-lines of any color ($0$, $1$, up to $D$) on $v$ and $\bar v$ will always be connected together via two-point graphs.
\end{itemize}
Therefore, for every such pair of vertices $v$ and $\bar v$ of $\cB$,
the two half-lines of color $0$ must be connected via some two-point graph in $\cG$ (see figure \ref{fig:factorization}). In other words, starting with a melonic $D$-colored observable, there is a \emph{unique} way to pair its external half-lines with two-point insertions so as to get melonic $(D+1)$-colored graphs. Then, the full expectation value is obtained by inserting this way full two-point functions, one for each pair of vertices which are joined.

\begin{figure}[htb]
\subfigure[Melonic observables.]{\includegraphics[width=6cm]{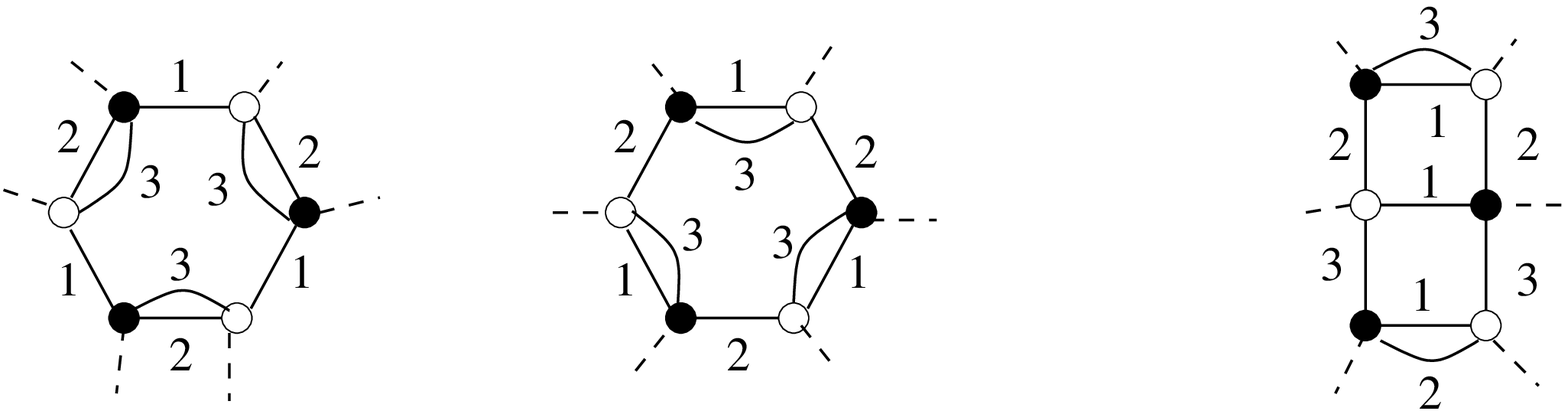} \label{fig:factorizationobs}}\hspace{2cm}
\subfigure[Graphs contributing to the melonic observables.]{\includegraphics[width=6cm]{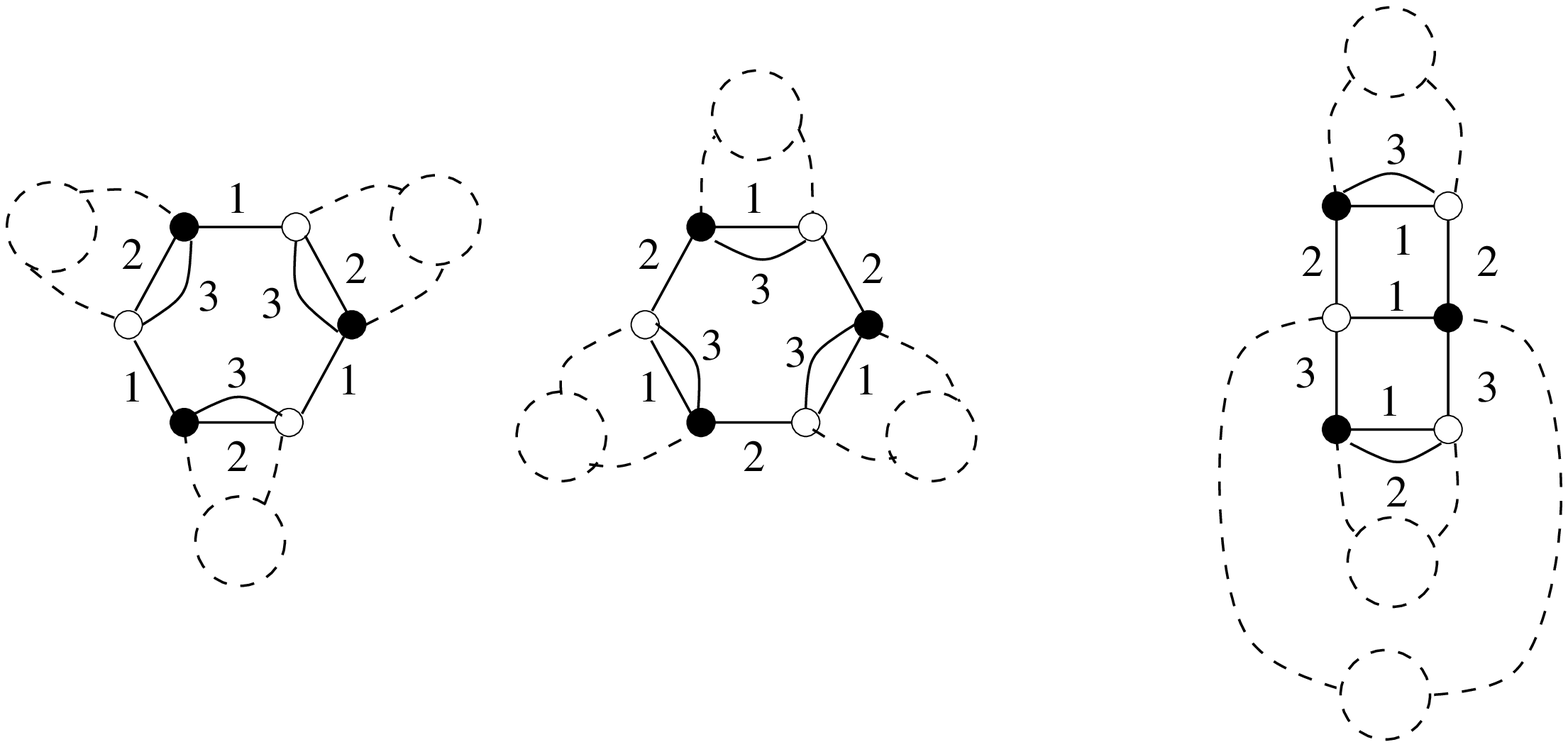} \label{fig:factorization}}
\end{figure}

As a result, in the $N\to \infty$ limit, the expectation of a melonic observable factors in terms of full two-point functions (dressed propagators).
The full two-point function writes
\bea
\frac{1}{Z} \Big{\langle} T_{n_1\dotsc n_D} \, \bar T_{\bar{n}_1\dotsc \bar{n}_D}    \Big{\rangle}
= \frac{\prod_{i=1}^D \delta_{n_i,\bar{n}_i}}{N^{D-1}}\ U(t_{\cB},N) \; , \qquad U(t_{\cB},N) = \frac{1}{t_1} + \dotsb \; ,
\eea
where $\frac{1}{t_1}$ is the bare propagator and the dots denote the radiative corrections.
We denote $\lim_{N\to \infty} U(t_{\cB},N) = U(t_{\cB})$. The large $N$ expectation of the
$D$-dipole observable $\cB_1$ computes then
\bea
 \lim_{N\to \infty} \frac{1}{N} \frac{\Big{\langle}  \Tr_{\cB_1}(T,\bar T) \Big{\rangle} }{Z} = U(t_{\cB}) \; .
\eea
Some graphs contributing to this expectation for $D=3$ are presented in figure \ref{fig:Ugraphs}, where the marked graph $\cB_1$ is presented
in bold. The colors of the lines are assigned turning clockwise $0$, $1$, $2$ and $3$ at the white vertices.

\begin{figure}[t]
\begin{center}
 \includegraphics[width=15cm]{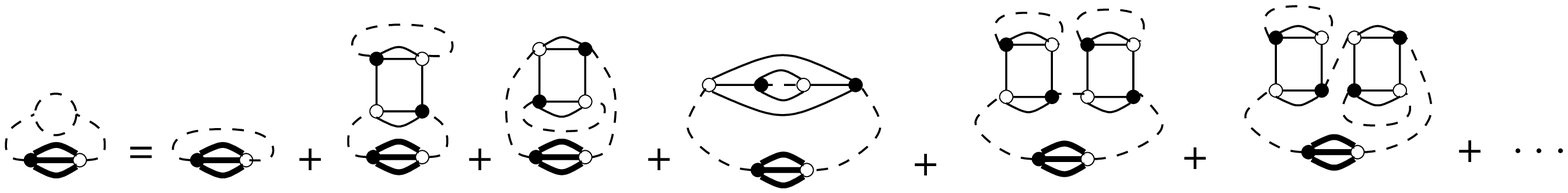}
\caption{Graphs contributing to the $3$-dipole expectation.}
\label{fig:Ugraphs}
\end{center}
\end{figure}

We thus conclude that
\be \label{universalfactorization}
\lim_{N\to \infty} \frac1N \frac{ \Big\langle\Tr_\cB(T,\bar T)\Big\rangle }{Z}=
  \begin{cases}
      0 \; , \qquad &\text{ if } \cB \text{ is not melonic} \\
      U(t_{\cB})^{p_{\cB}} \; , \qquad &\text{ if } \cB \text{ is melonic with } 2p_{\cB} \text{ vertices}
  \end{cases} \; .
\ee
In particular this factorization holds for the Gaussian model with $t_{\cB}=0$ for $\cB\neq \cB_1$, and the full two-point function
simply given by the bare covariance $U(t_{\cB})=1/t_1$.

The universality of tensor measures, first derived in \cite{Gurau:2011kk} (where it was obtained by mapping melons to trees), is the fact that the observables satisfy \eqref{universalfactorization}. It means that in the large $N$ limit the models become Gaussian. However this large $N$ limit is very non-trivial, as the covariance of the large $N$ Gaussian is the {\it full}, resummed, two-point function. The rest of this paper is dedicated to explore the various multicritical behaviors and continuum limits governed by this resummed covariance.

\subsection{The leading order two-point function and free energy}

The full two-point function at large $N$ is determined by a self-consistency equation provided by a Schwinger-Dyson equation supplemented with the above factorization. The relevant Schwinger-Dyson equation is
\be \label{SD1}
\frac1{N^D}\sum_{n_1,\dotsc,n_D} \frac1Z\ \int dT d\bar T\ \frac{\partial \phantom{T}}{\partial T_{n_1\dotsc n_D}}
\Bigl[T_{n_1\dotsc n_D}\, e^{ -N^{D-1} S(T,\bar T)}  \Bigr] = 0\;.
\ee
Taking the derivative explicitly, one gets
\be
1- \sum_\cB p_{\cB}\, t_\cB\; \frac1N \frac{\Big\langle  \Tr_\cB (T,\bar T)\Big\rangle}{Z} = 0\;,
\ee
where $p_{\cB}$ denotes the half-number of vertices of the bubble $\cB$ (that is the number of either black
or white vertices). At leading order in $1/N$ this can be rewritten in the following form. We first define the
leading order potential
\bea
 V(x,t_\cB) = \sum_{n\ge 1}  \Bigl(  \sum_{\substack{\cB \text{ melonic},\\ p_{\cB} = n }} t_\cB \Bigr) \;  x^n
\; ,\qquad   V'(x,t_\cB) \equiv \frac{\partial V}{\partial x}(x,t_\cB) = \sum_{n\ge 1} n \; \Bigl(  \sum_{\substack{\cB \text{ melonic},\\ p_{\cB} = n }} t_\cB \Bigr) \;  x^{n-1}  \;,
\eea
and taking into account the factorization of the melonic expectations, the Schwinger-Dyson equation becomes the following self-consistency equation
\bea \label{melonsresum}
  U(t_{\cB}) V'\bigr( U(t_{\cB}),t_\cB \bigl) =1 \; .
\eea
The leading order two-point function is the solution of this polynomial equation whose coefficients are the coupling constants of melonic observables.

Once $U$ is determined using \eqref{melonsresum}, one can access the free energy $f_0$. The leading order free energy $f_0(t_{\cB})$, like the leading order potential $V(x,t_\cB) = \sum_{n\ge 1}  \Bigl(  \sum_{\substack{\cB \text{ melonic},\\ p_{\cB} = n }} t_\cB \Bigr)  x^n $ and the leading order two-point function $U(t_{\cB})$ only depends on the coupling constants of the melonic bubbles $t_{\cB}$. Consider the function $f_0 - V  \bigl( U(t_{\cB}),t_\cB \bigr)  + \ln U(t_{\cB})  $. Its differential is
\bea
d \Bigl[ f_0 - V  \bigl( U(t_{\cB}), t_\cB \bigr) + \ln U(t_{\cB}) \bigr]
= \sum_{\cB \text{ melonic}} \Bigl[ \frac{\partial f_0 }{ \partial t_{\cB}} - U(t_{\cB})^{p_{\cB} }
   -  V'\bigl( U(t_{\cB}),t_\cB \bigr) \frac{\partial U}{\partial t_{\cB}} + \frac{1}{U(t_{\cB}) } \frac{\partial U}{\partial t_{\cB}}
\Bigr] \; dt_{\cB} =0 \;.
\eea
Thus the leading order free energy is
\bea\label{eq:freeenergy}
 f_0(t_\cB) = V  \bigl( U(t_{\cB}), t_\cB \bigr)  - \ln U(t_{\cB}) \; .
\eea

\subsection{The continuum limit}

Tensor models are of combinatorial nature and as such provide a notion of continuum limit in a combinatorial way. The idea
is that disregarding the geometrical content and interpretation which may be given to a model, this limit is always obtained as
the regime where graphs with a very large number of vertices dominate. As an illustration, we derive the continuum limit in
a particular $T^4$ truncation defined by the action
\be
S_{T^4}(T,\bar T) = \sum_{\vec n} T_{\vec n} \bar T_{\vec n} + g \sum_{\substack{a_1,\dotsc,a_D \\b_1,\dotsc,b_D}}
T_{a_1\dotsb a_{D-1} a_D}\,\bar T_{a_1\dotsb a_{D-1} b_D}\, T_{b_1\dotsb b_{D-1} b_D}\,\bar T_{b_1\dotsb b_{D-1} a_D} \; .
\ee
Note that the interaction term is melonic. The leading order potential is defined by $t_1=1, t_2=g$, that is $V(x) = x + gx^2$.
The leading order free energy is therefore
\be
f_0(g) = \sum_{n\in \N} g^{n} f^{(4n)}\;,
\ee
where $f^{(4n)}$ is the number of $(D+1)$-colored melonic graphs built with $n$ effective interactions $T^4$ (thus having $4n$ black and white vertices).
The number $f^{(4n)}$ is a canonical partition function for graphs with fixed number of vertices and $f_0(g)$ is its associated
grand-canonical partition function with lattice ``chemical potential'' $g$. The thermodynamic limit is encoded into the asymptotic behavior of $f^{(4n)}$,
\be
f^{(4n)} \underset{n \to \infty}{\sim} A\ n^{\gamma-3}\ g_c^{-n}\;,
\ee
for some constants $A$, $g_c$ and $\gamma$. Thus $g_c$ is the radius of convergence of $f_0(g)$, which means that when $g$ approaches $g_c$,
$f_0(g)$ loses its summability and graphs with a large number of vertices ($4n$) dominate its behavior.
The power-law decay characterized by $\gamma$ controls the singularity of $f_0(g)$ close to $g_c$, since
\be
f_0(g) \sim \vert g-g_c\vert^{2-\gamma}\;.
\ee
The exponent $\gamma$ is known as the entropy exponent\footnote{In the $\Tr M^4$ matrix model for random two-dimensional lattices for instance one has $\gamma=-1/2$ \cite{Di Francesco:1993nw}, which is the universality class of pure 2d quantum gravity. As such, it is reached generically, i.e. for most values of the coupling constants, in the continuum limit of one-matrix models.}.
Let us compute the entropy exponent of the model defined by $S_{T^4}$.
First one notices that the derivative of the leading order free energy writes in terms of the leading order two-point function
$\frac{\partial f_0}{\partial g}=U(g)^2$. The equation \eqref{melonsresum} gives
\be
2g\, U(g)^2 +U(g) -1=0\;,\quad \text{hence}\quad U(g) = \frac{\sqrt{1+8g} -1}{4g}\;,
\ee
where we selected the physical root with initial condition $U(0)=1$. One thus identifies $g_c=-1/8$ and for $g\to g_c $ the non-analytic parts
of the two-point function and free energy are
\be
U(g)_{\rm sing}\sim (g-g_c)^{\frac{1}{2}}\;,\qquad f_{0, {\rm sing}} \sim (g-g_c)^{3/2}\;,\qquad \text{hence $\gamma=1/2$}\;.
\ee
The average number of effective interactions (proportional to the number of vertices) diverges when tuning to criticality
\be
\langle n\rangle \,=\, g\,\frac{\partial}{\partial g} \log f_0 \,\sim\, \frac{1}{\vert g-g_c\vert}\ \underset{g\to g_c}{\to}\ \infty \;.
\ee

\subsection{Virasoro constraints}

We now come back to the generic model with arbitrary couplings. The full set of Schwinger-Dyson equations is obtained by
inserting generic $D$-bubble observables in \eqref{SD1} . The equations can be recast as $L_\cB Z(t_{\cB}) =0$ for some differential
operators $L_\cB$ labeled by the observables. The algebra of these operators has been discussed at length in \cite{Gurau:2011kk}.
Due to the large $N$ factorization one can find the leading order Schwinger Dyson equations and the associated algebra of constraints
by a shorter route. We show below that the large $N$ factorization \eqref{universalfactorization} reduces the Schwinger-Dyson equations
to a set of Virasoro constraints, like in matrix models. We emphasize that this only holds at leading order in $1/N$.

Note that in fact the leading order two-point function $U(t_{\cB})$ as well as the leading order free energy $f_0$
depend only on the sums of the coupling constants of melonic observables at fixed number of vertices. Thus, defining $t_n \equiv \sum_{\substack{\cB \text{ melonic},\\ p_{\cB} = n } }  t_\cB$, the large $N$ factorization becomes
\be
\frac{\partial f_0}{\partial t_n}=[U(t_p)]^n\;,
\ee
Then multiplying \eqref{melonsresum} by $U^k$ for any positive $k$ we get
\be
[U(t_p)]^k-\sum_{n\geq 1} n\, t_n\,[U(t_p)]^{n+k} = 0 \;, \quad  \forall k\geq0 \; .
\ee
These equations can be seen as differential equations on $f_0$,
\be
L_k\,f_0=0\;,\quad \text{for}\quad L_k = \frac{\partial }{\partial t_k} - \sum_{n\geq 1} n\,t_n\,\frac{\partial}{\partial t_{n+k}}\;.
\ee
It is straightforward to check that the differential operators satisfy the well-known algebra
\be
[L_n, L_m] = (m-n)\,L_{m+n}\;.
\ee

\section{Multicritical behaviors in the large N limit}\label{sec:multic}

We now set the summed coupling constants $t_1 = 1/g$, and $t_n = \alpha_n/g$ to investigate the different possible critical behaviors
with respect to the parameter $g$. The potential $V$ becomes $V(x)=(x+\sum_{k=2} \alpha_k\, x^k)/g$. The self-consistency equation
\eqref{melonsresum} for the two-point function becomes
\be \label{eq:t1}
g = U + \sum_{k=2}^m k\, \alpha_k\, U^k \;.
\ee
Like in the matrix models review \cite{Di Francesco:1993nw}, the continuum limit is obtained by setting $\partial g/\partial U=0$. At least locally, this equation can be solved for $U$ in terms of the parameters $\alpha_k$. One concludes that for generic values of these parameters $\gamma=1/2$, thus proving the universality of this continuum limit, first derived in \cite{Bonzom:2011zz} for the colored model.

But there are points on the set of parameters where the equation $\partial g/\partial U=0$ can not be solved for $U$, because $\partial^2 g/\partial U^2$ may vanish and there the implicit function theorem can not be applied. In these cases, one has multicritical behaviors and $\gamma>1/2$. A multicritical point of order $m$ is defined, like in one-matrix models, by
\be
\frac{\partial g}{\partial U} = \dotsm = \frac{\partial^{m-1} g}{\partial U^{m-1}}  = 0  \;,
\qquad \text{and}\qquad  \frac{\partial^{m} g}{\partial U^{m}} \neq 0 \;,
\ee
which imply $g= g_c - (U_c-U)^m + \mathcal{O}((U-U_c)^{m+1})$. Such multicritical behaviors have already been observed in tensor models
in \cite{Bonzom:2012sz} (where they are interpreted in terms of dimer models) and \cite{doubletens}. As $g$ is a polynomial
in $U$ whose coefficients $\alpha_k$ can be freely chosen, multicritical points can be reached for the generic one-tensor model.
We present below a minimal realization, i.e. a potential $V$ with minimal degree leading to a multicritical
point of order $m$. We first set $U_c = m^{-\frac{1}{m-1}}$, and consider
\be
V(U) = \frac{1}{g} \sum_{k=1}^{m} \frac1{k}\,U_c^{m-k}\, \Bigl[U_c^k - \bigl(U_c-U\bigr)^{k}\Bigr]\;.
\ee
Thus $V$ is of degree $m$, satisfies $V(0)=0$, the coefficient of the linear term is $1/g$ and $UV'(U)= [U_c^m-(U_c-U)^m]/g$.
The self-consistency equation is exactly
\be
g = g_c - (U_c-U)^m \;, \qquad \text{with} \qquad g_c = U_c^m \;.
\ee
Substituting into \eqref{eq:freeenergy} we find
\bea
  f_0 &=& \frac{1}{g}  \sum_{k=1}^{m} \frac1{k}\,U_c^{m-k}\, \Bigl[U_c^k - \bigl(U_c-U\bigr)^{k}\Bigr]\;- \ln \bigl[ U_c - (U_c-U)\bigr] \crcr
   &=& f_0(g_c) - \frac{U_c^m}{g} \sum_{k=1}^{m} \frac1{k}\, \biggl( \frac{U_c-U}{U_c} \biggr)^k
   + \sum_{k=1}^{\infty} \frac1{k}\, \biggl( \frac{U_c-U}{U_c} \biggr)^k 
   +\biggl[\frac{1}{g}-\frac{1}{g_c}\biggr] \sum_{k=1}^{m} \frac1{k}\,U_c^m\; .
\eea
Taking into account that $ U_c -U = (g_c-g)^{\frac{1}{m}} $, we obtain for $g\to g_c$
\bea
  f_0 = f_0(g_c) + \biggl(\sum_{k=1}^m\frac1k\biggr) \biggl(\frac{g_c-g}{g_c}\biggr)
-\frac{m}{(m+1)} \biggl(\frac{g_c-g}{g_c}\biggr)^{1+\frac{1}{m}} +\ \mathcal{O}\bigl( (g_c-g)^{1+\frac{2}{m}} \bigr) \; ,
\eea
hence a multicritical entropy exponent $\gamma_m=1-\frac1{m}$, as obtained in \cite{doubletens} and in \cite{Bonzom:2012sz},
coinciding with the ones of multicritical branched polymers \cite{BP-ambjorn}.


\section*{Acknowledgements}
Research at Perimeter Institute is supported by the Government of Canada through Industry
Canada and by the Province of Ontario through the Ministry of Research and Innovation.

\appendix

\renewcommand{\theequation}{\Alph{section}.\arabic{equation}}

\section{Combinatorics of colored graphs: jackets, degree and melons} \label{app}

In order to define the degree of a graph one first needs the notion of \emph{jacket} \cite{Gur3,GurRiv,Gur4}.
\begin{definition}
Let $\cB$ be a $D$-colored graph and $\tau$ be a cycle on $\{1,\dotsc,D\}$. A colored {\bf jacket} $\cJ$ of $\cB$ is a ribbon graph
having all the vertices and all the lines of $\cB$, but only the faces with colors $(\tau^q(1),\tau^{q+1}(1))$, for $q=0,\dotsc,D-1$,
modulo the orientation of the cycle.
\end{definition}
As a jacket $\cJ$ of $\cB$ contains all the vertices and all the lines of $\cB$, $\cJ$ and $\cB$ have the same
connectivity. As such, any jacket $\cJ$ carries some key topological information
about $\cB$ (for instance the fundamental group of $\cB$ is a subgroup of the fundamental group of any
of its jackets \cite{BS3}). For graphs with four colors, the jackets correspond to Heegaard splitting surfaces \cite{Ryan:2011qm}.

Jackets are ribbon graphs, hence they are completely classified by their genus $g_\cJ$.
\begin{definition}
The {\bf degree} $\omega(\cB)$ of a colored graph $\cB$ is the sum of genera of its jackets, $\omega(\cB)=\sum_{\cJ} g_{\cJ}$.
\end{definition}

Graphs with three colors are ribbon graphs, and the degree coincides with the genus.
The crucial property of the degree is that the total number of faces $\cF$ of a $D$-colored graph computes in terms of the degree.

\begin{proposition}\label{prop:crucial}
Let $\cB$ a $D$-colored graph with $2p$ vertices. Then the total number of faces of $\cB$ respects
\bea
 |\cF| = \frac{(D-1)(D-2)}{2}p + (D-1) - \frac{2}{(D-2)!} \omega(\cB)\; .
\eea
\end{proposition}
{\bf Proof:} This equation can be found in the literature (see \cite{Gur4} for instance). However, due to its importance we present here its proof.
Every jacket is a ribbon graph with $2p$ vertices and $D p$ lines, hence the number of faces of a jacket is
\bea\label{eq:int}
 |\cF_{\cJ}| = (D-2)p + 2 - 2g_{\cJ} \;.
\eea
As jackets correspond to cycles over $D$ elements modulo the orientation, $\cB$ has $\frac{1}{2}(D-1)!$ distinct jackets.
The faces with colors $ij$ will belong to the $2(D-2)!$ jackets corresponding to the cycles $\pi$ such that $\pi(i)=j$ and $\pi(j)=i$.
Moding by the orientation we conclude that each face belongs to exactly $(D-2)!$ distinct jackets. Summing \eqref{eq:int} over the
jackets and dividing by $\frac{1}{2}(D-1)! $ proves the lemma.

\qed

Of course the same definition goes trough for graphs $\cG$ with $D+1$ colors. Further facts concerning the degree are listed below.

\begin{proposition} \label{deg>deg}
The degrees of a $(D+1)$-colored graph $\cG$ and of its $D$-bubbles $\cB_{(\rho)}$ with colors $1,\dotsc, D$ respect
\bea
  \omega(\cG) \ge D \sum_{\rho} \omega(\cB_{(\rho)})\; .
\eea
\end{proposition}
The proof of this statement can be found in \cite{Gurau:2011tj}, lemma 7. The proof relies on the identification of the jackets of $\cB_{(\rho)}$
as ribbon subgraphs of the jackets of $\cG$.

\begin{proposition} \label{prop:planarjacket}
If the degree of a $(D+1)$-colored graph $\cG$ vanishes, i.e. all its jackets are planar, then $\cG$ is dual to a $D$-sphere.
\end{proposition}

The proof of this lemma can be found in \cite{Gur4}.

The graphs of degree $0$ have been thoroughly analyzed in \cite{Bonzom:2011zz}. Their characterization relies on two lemmas.

\begin{proposition} \label{prop:face-2vertices}
If $D \geq 3$ and $\cG$ is a $(D+1)$-colored graph with vanishing degree, then $\cG$ has a face with exactly two vertices.
\end{proposition}
{\bf Proof:} All faces of $\cG$ have an even number of vertices. Denote $|\cF_s|$ the number of faces with $2s$ vertices.
By proposition \ref{prop:crucial} (taking into account that $\cG$ has $D+1$ colors), the total number of faces of
$\cG$ is $|\cF| =\sum_{s\ge 1} |\cF_s|= \frac{D(D-1)}{2}p +D$.
As a vertex belongs to
$\frac{D(D+1)}{2}$ faces we have $\sum_{s\ge 1} s |\cF_s| = \frac{D(D+1)}{2}p$. Eliminating $\cF_2$ we get
\be
\cF_1 = 2 D + \sum_{s\ge 3} (s-2) \cF_{s} +  \frac{D(D-3)}{2}\, p \; .
\ee
The first two terms give a strictly positive contribution for any $D$, whereas the third term changes sign for $D = 3$.
Thus $\cF_1 \geq 1$ only for $D\geq3$.

\qed

The fact that this proposition fails in $D=2$ is the source of the difference between the large $N$ limit of matrix models, dominated by planar graphs,
and the large $N$ limit of tensors of rank $D$, dominated by {\it melonic} graphs, which we describe below.

\begin{proposition} \label{prop:melon}
If $D \geq 3$ and $\cG$ is a $(D+1)$-colored graph of vanishing degree, then it contains two vertices $v$ and $\bar v$ separated by
exactly $D$ lines.
\end{proposition}

The proof of this lemma can be found in \cite{Bonzom:2011zz}.
It relies on proposition \ref{prop:face-2vertices}, but as it is somewhat convoluted we do not reproduce it here.

We exploit this lemma in the following way. Starting from a $(D+1)$-colored graph $\cG$ we identify two vertices
$v$ and $\bar v$ separated by $D$ lines. Erasing this subgraph and reconnecting its external lines we obtain a
graph having two less vertices and degree $0$ (as it can be checked explicitly). Iterating this erasing procedure
we necessarily end up with a $(D+1)$-dipole, that is the graph having two vertices connected by $D+1$ lines.
Conversely, every melonic graph can then be obtained by starting from the $(D+1)$-dipole and inserting
such subgraphs (consisting in two vertices connected by $D$ lines) arbitrarily on all the lines.


\end{document}